# Causal Organization Prior to and Promoting Self-Replication in a Catalytic Model of the Origin of Life

F. Pigozzi[1], M. Levin[1,2]*

**Affiliations:**

[1]Allen Discovery Center, Tufts University; Medford, 02155, USA.

[2]Wyss Institute for Biologically Inspired Engineering, Harvard University; Boston, 02115, USA.

*Corresponding author. Email: michael.levin@allencenter.tufts.edu

**Abstract:**

Agents that exert causal power in the world are thought to be the product of selection among diverse replicators. Whether organized causal structure can arise before replication and influence its emergence remains unclear. Here we examine this question in the Graded Autocatalysis Replication Domain (GARD), a model of catalytic molecular assemblies capable of compositional self-reproduction. We find that causal emergence—irreducible predictive information generated by the system as a whole—exhibits punctuated dynamics associated with self-replication. Early causal emergence predicts subsequent self-replication more accurately than established measures of molecular dynamics. Moreover, manipulating molecular assemblies to increase causal emergence raises the persistence and abundance of self-replication, whereas the opposite is true for interventions that decrease causal emergence, indicating a functional control knob. These results show that causal organization can precede the appearance of replicators and can influence their subsequent dynamics. They suggest that increases in integrated causality can shape the organization of active media before evolutionary dynamics begin.

## Introduction

A key aspect of intelligence and agency is the ability to exert causal power on the physical world. While debates about downward and circular causation continue [1-9], central to our understanding of cognition is the ability for a composite system, such as a living organism, to exhibit a kind of integration that makes it a driver of subsequent events [10, 11]. Recent work in causal information theory has significantly advanced our ability to quantify when and how much a given system is an integrated causal agent [12-22]. Consider the anecdotal example of a swarm of ants: if all are from the same colony, they will have a higher causal integration because of their ability to communicate through pheromones, while a swarm of ants from different colonies will likely have lower causal integration. These tools are also used in clinical neuroscience, for example, to detect human agency in the brain [23].

The conventional paradigm focuses on natural selection as the mechanism that produces strongly integrated systems with high causal power. The fundamental engine, on this view, of the remarkable intelligence we observe in the living world is selection: differential success of reproduction is thought to power the rise in agency across the tree of life. Thus, the ability of evolution to ratchet up causal power is predicated on the existence of replicators. What happens before there are replicators? What does the causal architecture of an active medium look like before replicators appear within it? The field of diverse intelligence seeks to test the applicability of tools from cognitive and behavioral science to contexts beyond brainy animals [24-29], identifying a continuum of scaling dynamics that underlie the journey from minimal active matter to conventionally intelligent agents. Causal information tools have already been applied to novel scenarios such as cancer [30], evolutionary dynamics [31], and even sports [32]. Might they be useful tools to study the origin of replication and the path to the existence of a minimal evolutionarily significant unit?

The emergence of persistent replicative systems depends not only on the replicators themselves but also on the physicochemical regimes that sustain them. Studies of thermodynamically driven chemical systems [33], autocatalytic networks [34-36], and growth-first scenarios [37] suggest that organized, self-maintaining dynamics can arise before the appearance of genetically encoded heredity and may provide the conditions that enable later evolutionary transitions. Against this backdrop, an important open question is whether the causal architecture of the pre-replicative medium not only anticipates but also influences the appearance and persistence of self-reproducing states.

This question is particularly important in the Origin of Life (OoL), where the goal is to study the transition to replicators within an active medium [9, 38-41]. This is the case with early Earth scenarios (abiotic to biotic transition) [36, 42, 43] as well as in silico [44-49]. Classical approaches have focused on the generation of life's building blocks and proto-metabolic networks, including the primordial soup [50, 51], RNA world [52], and metabolism-first [40, 53] scenarios. More recent work has emphasized self-organization, information processing, and causal structure, viewing the abiotic-to-biotic transition as the onset of systems capable of storing, transmitting, and acting upon information. Here, we used information theory tools to study the causal properties of an active medium during the *de novo* emergence of replicators. We chose the popular Graded Autocatalysis Replication Domain (GARD) model [54-64] - a widely used simulation-tractable framework that captures several dynamics central to life's origin, including the emergence of mutually catalytic molecular assemblies and compositional (rather than sequence-based) inheritance. We studied compositional dynamics that allow the appearance of self-reproducing catalytic networks that are considered relevant to hypotheses about the origin of life on Earth. GARD simulates assemblies that grow by accretion of environmental molecules. These assemblies have been found to self-replicate, settling into recurring molecular compositions that exhibit homeostatic growth. We asked whether detectable information dynamics could predict or even control aspects of replicator formation.

## Results

We simulated 100 independent GARD assemblies (with different random seeds) that grew by random accretion of different types of molecules (see Materials and Methods). In Figure 1, we summarized their evolution. Assemblies underwent fission when they reached a critical size, and the simulation continued from one of the daughter assemblies. Simulations stopped after a fixed number of generations. To capture the conditions of the prebiotic soup, the GARD model is essentially a stochastic model of events before natural selection takes place. Self-replicators emerged spontaneously (as a product of specific GARD dynamics and parameters) as recurring compositions that are "inherited" across generations [64, 65]. In other words, a given simulation can be either in a self-replicator or not ("drift") state at any given time step. Even if self-replicators emerged consistently in our simulations, an assembly could still enter and exit self-replication during its lifetime, depending on whether it crosses a given similarity threshold relative to the most recurring composition [65].

To uncover links between causal information theory and this bio-realistic OoL hypothesis, we computed our causal emergence measure $\Phi^r$ of the entire soup for each run, using the molecular type counts of the simulated assembly as substrate (see Materials and Methods). To ensure $\Phi$ captured information that is not already distilled by prior existing metrics, we computed the correlation (both Spearman's and Pearson's) between $\Phi^r$ trajectories and several established network properties (number of nodes and edges, in-degree, out-degree, betweenness centrality, PageRank, HITS scores) of the catalytic graph and dynamical properties (sample entropy, correlation dimension, Lyapunov exponents, DFA, generalized Hurst exponent) of the simulations. We found no significant correlations across all comparisons, indicating that $\Phi$ captured an original axis of self-organization not apparent in other metrics.

Initially, we aggregated $\Phi^r$ over the 100 runs and found no significant trend over molecular time (linear regression, p=0.1995) at the large scale, as shown in Figure 2A. However, most runs exhibited punctuated, spiking events in $\Phi^r$ (Figure 2B, 2C, 2D) that were more than three standard deviations above the overall mean (even if not monotonically increasing), indicating that, despite the lack of an aggregate trend, nontrivial information flows occurred within the molecular substrates when examined at the individual-simulation level.

We hypothesized that these spiking events could be linked to the emergence of self-replicators and tested this correlation using the Spearman correlation test (Figure 3). For each run, correlation was computed using the entire trajectory. A majority (73/100) of runs showed a positive correlation between $\Phi^r$ and whether self-replication was present or not, with a majority of these (54/73) being significant. In other words, as self-replication took place, so $\Phi^r$ increased. Indeed, $\Phi^r$ was significantly higher (using Mann-Whitney, $p<0.001$) for assemblies during time steps that belonged to a self-replicator than those that did not in most runs (57/100). In Figure 4, we show the causal emergence for individual runs during self-replication (4A) and with the median ± std (4B). Using Fisher's method (p<0.001), the combined evidence from the 100 independent runs significantly supported that causal emergence was higher during steps when the assemblies were in self-replication, highlighting that self-replication corresponded to an increase in causal emergence of the prebiotic chemistries. Indeed, the null hypothesis of temporal independence was rejected in 86/100 runs (median Ljung-Box $p$=2.07x10$^{-51}$), indicating that $\Phi^r$ trajectories possess statistically significant temporal structure rather than behaving as independent samples. Even after differencing, 100/100 runs retained significance, indicating that the observed temporal structure cannot be explained by drifts or trends.

Having verified this generally positive correlation between self-replication and causal emergence, we tested whether either was temporally predictive of the other: do rises in $\Phi^r$ precede and predict the appearance of replicators? To do so, we recomputed $\Phi^r$ on the first 25% points of each run and asked a machine learning model to predict the remaining 75% of the self-replication trajectory from the same GARD run. As a machine learning model, we fitted an MLP, using 80% of the runs for training and 20% for testing (bootstrapped across 10 repetitions with different seeds). To exclude confounds, we trained the same model using other soup statistics as input (change in composition, raw composition counts, and fluxes between molecular species) and a dummy baseline that always predicted the most abundant label (self-replication or not). Figure 5 shows the results, where each dot in the scatterplots represents one of the 10 repetitions. The model trained on $\Phi^r$ beat all other baselines (all differences significant with Mann-Whitney, $p<0.01$), showing it could fit a non-linear decision boundary to predict when a GARD assembly self-replicated. We repeated the experiment with different input/output split proportions and found quantitatively similar results. Thus, the spikes in $\Phi^r$ carried information predictive of when future self-replication would occur. Indeed, we found a positive and significant Spearman's correlation between the self-replication probability and the distance among spikes ($\rho$=0.71, $p<0.001$). Conversely, the correlation with the $\Phi^r$("height") of the spike was not significant. Altogether, these results show that causal emergence is a viable candidate for predicting the early emergence of self-replication in prebiotic soups.

However important for applications, predictive ability does not prove causality: Is causal emergence a *driver* of self-replication? Was $\Phi^r$ not just a detector of impending replicators, but also functionally important for their appearance? We carried out the following gain-of-function experiment, illustrated in Figure 6A: we simulated the same 100 assemblies as before, but at each generation, we asked which interventions would raise its $\Phi^r$. To answer this, we searched all possible additions and deletions of molecules (there were twice as many molecule types, 100, so an exhaustive search was computationally tractable) and applied the one that raised $\Phi^r$ the most. These interventions occurred after every fission event and were followed by the simulation of the assembly using the usual GARD dynamics. To assess causality, we measured the impact of these interventions on self-replication and compared the results with minimization (instead of maximization) of $\Phi^r$ and with no intervention (control, i.e., the original runs).

The results are summarized in Table 1 according to four properties of self-replicators in GARD assemblies: self-replicator persistence, probability, consistency (measured as the Pearson's $\rho$ between consecutive steps; the higher the value, the more compact and less erratic the self-replicators), and time to the appearance of the first replicator. Interventions that drove $\Phi^r$ up increased the persistence and consistency of self-replicators ($p<0.001$, Mann-Whitney), indicating more organized self-replication. Figure 6B corroborates this finding with the distribution of persistence by treatment. While the overall probability of self-replication did not significantly differ between max$\Phi^r$ and control, we observed a consistent increase over simulation time for the max$\Phi^r$ (Figure 6C), consistent with interventions that slowly drove the self-replication probability up as they accumulated. This linear trend was statistically significant ($p<0.001$, n.s. for the control) and led max$\Phi^r$ simulations to achieve higher self-replication probability by the last GARD generation. Conversely, the min$\Phi^r$ interventions led to a significant worsening ($p<0.001$, Mann-Whitey) in all four properties, showing that $\Phi^r$-directed interventions causally modulated self-replication.

These results highlighted how $\Phi^r$ is not only be a signature of the early appearance of self-replication but also provided a control knob for ushering in (or staving off) self-replicators: driving causal emergence up.

## Discussion

A key feature of living systems is that they initiate actions in the world – they have causal potency, if anything does [66-68], and organize their components toward an integrated collective that is "more than the sum of its parts" – a property which the recent tools of information theory seek to quantify [14, 69-72]. Origin-of-life research, like embryology, seeks to zoom in on the continuum between passive matter and active agents, to understand how agents arise from passive media. Once replicators appear, Darwinian dynamics guarantee their further evolution via the selection of more successful variants. But what happens *before* there is a persistent, tangible object that replicates with differential success? We used the GARD model to examine the causal architecture of a soup of interacting molecules, which has been used to study the dynamics of the prebiotic-to-biotic transition.

Our analysis indicated that spikes in causal emergence were associated not only with the appearance of self-replicators, but also with the initial appearance of replicators later in the simulations. Most $\Phi^r$ trajectories rejected the null hypothesis of white noise, indicating they retain temporal memory; in other words, the spiking events of $\Phi^r$ leading up to the appearance of replicators were not independent but reflected a progressive process. Crucially, intervening on the assemblies to maximize their $\Phi^r$ yielded a higher self-replication probability than the original runs, whereas intervening to minimize it reduced self-replication probability. Overall, these data reveal interesting dynamics that presage and predict the appearance of replicators, consistent with the idea that organizational and information-first dynamics emerge alongside the growth processes emphasized in metabolism-first accounts [37]. Since $\Phi^r$ is agnostic to the material implementation, it is compatible with many different media but emphasizes key dynamics that are not intrinsically tied either to energetics or a genetic code. Interesting aspects for further evaluation include the fact that the $\Phi^r$ does not need to remain high for replicators to appear, just to spike, suggesting that the substrate (the molecular compositions upon which $\Phi^r$ is computed) carries the memory of the spike imprinted on it. Because our analyses concern compositional self-reproduction prior to persistent genetic heredity, they do not address later evolutionary challenges such as the error-threshold and parasite problems associated with template-based replicators [73]. These later transitions may require cooperative or multilevel organizational mechanisms beyond those modeled here [74].

It should be noted that in addition to the relevance of these results for the origins of life (on Earth and perhaps elsewhere), they may have implications for a wide range of dynamical systems. Our ongoing work seeks to characterize causal emergence across a variety of living, non-living, and hybrid systems at all scales, and to understand how and when causal emergence metrics can indicate the impending arrival of novel, often unexpected features in these systems. Future research on patterns in this context will be needed to extend our method and enable predictions of significant shifts ranging from sub-molecular to the ecosystem/society levels. Moreover, there are many real-world scenarios in which the appearance of replicators is to be hastened, and others in which it is to be avoided. These include technological contexts, such as nanomaterials and information networks, and biomedical settings where persistent physiological states are only beginning to be addressed as causal factors in health and disease [75]. Our data suggest that, in addition to prediction, manual fine-tuning of $\Phi^r$ is a useful method for managing the impending emergence of replicators. Moreover, quantifiable causal dynamics that presage the emergence of integrated systems before those systems exist raise interesting questions about the complementary roles of evolution and ordering dynamics inherent in mathematical properties that are not themselves the results of selection [34, 44, 76].

## Materials and Methods

### *GARD model*

The Graded Autocatalysis Replication Domain (GARD) [65] simulated particle assemblies that grew by the accretion of environmental molecules, emulating the prebiotic soup. Each molecule belonged to one of $N_g$ molecular types (the colors in Figure 1); these did not necessarily map to real-world molecular species but served as an ideal substrate.

We selected GARD because it provides a well-established framework for studying compositional reproduction in catalytic networks and has been proposed as a bridge between "metabolism-first" and other OoL scenarios [77, 78], such as the "RNA-first" hypothesis [79]. Moreover, GARD most closely matched the chemical networks from our preliminary studies of gene regulatory pathways [80-82]. The GARD model has undergone several iterations [78, 83]; we adopted the original for parsimony [65].

Each simulation traced the fate of a single assembly with a catalytic matrix β (Figure 1A, bottom), starting with $n_{min}$ molecules. The rates of the catalytic matrix $\beta$ were sampled from a lognormal distribution with mean $A$ and standard deviation σ since prebiotic systems likely had many weak catalysts and a few strong ones, while the initial molecular types were sampled uniformly without replacement from the $N_g$ possible types.

We iterated a simulation for $n_{gen}$ generations. At each generation, the assembly's composition underwent a sequence of stochastic (Poisson) updates that specified how many molecules of each type left the assembly for the environment and how many joined from the environment. These updates continued until the assembly reached the maximum size $n_{max}$ or the maximum number of steps $max_{steps}$. At the end of the generation, the assembly fissioned into two daughter assemblies (Figure 1B), whose constituent molecules were sampled from a binomial distribution with parameter 0.5.

Following the best results found in [65], we set $N_g$=100, $n_{min}$=40, $A$=-4, $\sigma$=4, $n_{gen}$=100, $n_{max}$=80, and $max_{steps}$=1000.

Under these parameters and dynamics, the GARD model settled on steady molecular compositions that were highly similar (in Euclidean space) from one generation to the next [65]. These recurring steady states were *self-replicators* [35], like attractors in dynamical systems [84]. They presented compositional heredity without genetic material, showing how life-like reproduction can emerge from purely chemical systems [85]. Because causal emergence requires a sufficiently dense trajectory to be reliably estimated, we used the full molecular trajectory from the simulations, not just the fission events. This means our self-replication tagged instantaneous homeostatic growth; when it occurred at fission events (as GARD expected), it became self-replication.

*Causal Emergence*

The GARD implementation was a system consisting of $N_g$ parts (the molecule type counts) that evolved over time *t*. Let

$$X_t^i$$

denote the expression level of molecule type *i* at time *t*. Following [86, 87], we say that the system is *causally emergent* if the whole system holds unique predictive information about its future evolution that cannot be derived from its parts. This unique predictive information can be estimated using the Partial Information Decomposition [88] and its recent extension to time series like ours, the ΦID decomposition [82, 89].

Other measures of integrated information exist, such as total correlation and co-information. But they are instantaneous measures; they fail to capture the temporal and causal aspects of information dependencies through all future time steps, a crucial feature for dynamical systems that evolve over time. The ΦID captures this dynamic and directional information flow over time and is thus more suited for our OoL simulations. Formally, causal emergence is:

$$\Phi^r = I(X_t, X_{t+1}) - \Sigma_i I\left(X_t^i, X_{t+1}\right)$$

Where

$$I(X_t, X_{t+1})$$

is the time-lagged multivariate Shannon mutual information. Estimating the multivariate mutual information above is computationally intractable for even moderate $N_g$, let alone the value we used. Hence, we adopted the minimum-information bipartition [90] to reduce the system to a computationally tractable two-component system. The minimum-information bipartition is defined so that the two components share the least information and hence have the weakest causal cut. Once a bipartition is computed, the $\Phi ID$ decomposes the total information of the system into information atoms, one of which is our causal emergence as expressed by $\Phi^r$ in the equation above.

Causal emergence is the information about the system's future that is generated jointly by the system as a whole—above and beyond what its parts independently or redundantly contribute—i.e., the system's irreducible causal influence across time.

*Data preprocessing*

Causal emergence requires an ideal substrate upon which to be computed. We used the relative molecular compositions of each assembly because they captured the full dynamical information of the system. Each simulation output a trajectory $X$ of size $N_g$ *x* $n_{tot}$, where $n_{tot}$ was the total number of molecular steps (different for each simulation due to stochasticity). Since relative molecular compositions are constrained to sum to 1, the data live on a simplex, creating spurious correlations. To solve this, we applied the centered log-ratio transform:

$$X_t = \log \frac{X_t}{geometric\ mean(\mathrm{X})}$$

By expressing each component as the deviation from the system-wide geometric mean (the only symmetric, scale-invariant reference), the centered log-ratio transform removes the closure constraint and makes the components comparable. It maps the simplex into a Euclidean space where the $\Phi ID$ assumptions hold, while preserving all relative information. The centered log-ratio transform solves closure, but it makes the components linearly dependent and thus renders the covariance matrix singular and the entropy estimates ill-defined. We therefore removed the last component to obtain our full-rank representation.

**Acknowledgments:** We thank Joana Xavier for helpful comments on the manuscript, and Patrick McMillen, Richard Watson, and numerous members of the Levin lab for useful discussions on this topic.

**Funding:** ML gratefully acknowledges support of Astonishing Labs, Templeton World Charity Foundation, Inc., and Grant 62212 from the John Templeton Foundation. The opinions expressed in this publication are those of the authors and do not necessarily reflect the views of the sponsors.

**Author contributions:**

Conceptualization: FP, ML

Methodology: FP

Investigation: FP, ML

Visualization: FP

Funding acquisition: ML

Project administration: ML

Supervision: ML

Writing – original draft: FP, ML

Writing – review & editing: FP, ML

**Competing interests:** This research was funded at Tufts under a Sponsored Research Agreement with Astonishing Labs. M.L. is a co-founder and shareholder of Astonishing Labs. Astonishing Labs has certain rights to any inventions associated with this research.

**Data, code, and materials availability:** All code to reproduce and replicate the experiments and results will be made publicly available upon publication.

**Supplementary Materials**

None.

## Figures

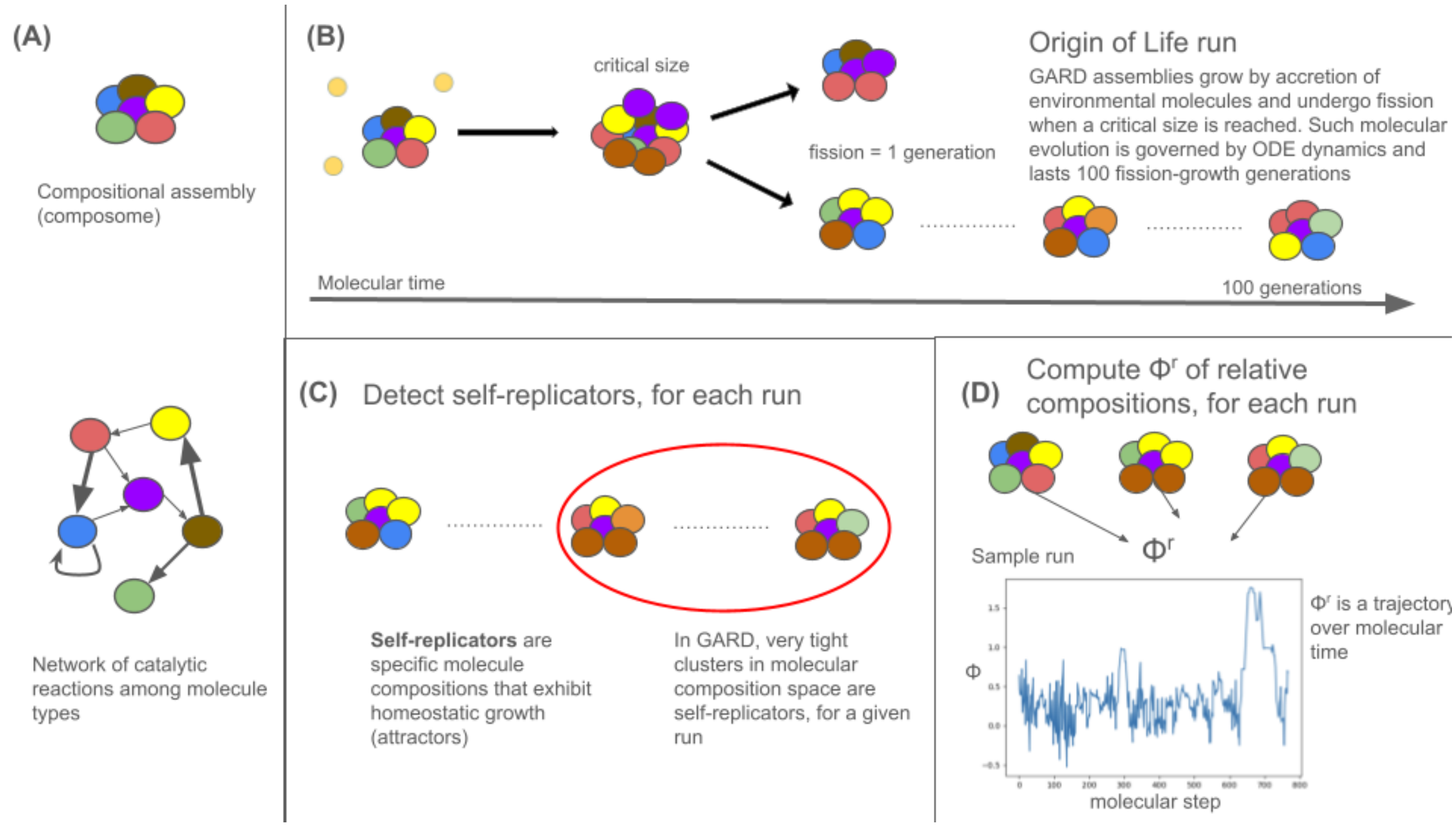

**Fig. 1. System for the study of causal emergence in prebiotic chemistry. A)**: The GARD model simulated compositional assemblies of molecules (the *composomes*, shown as colored circles, with one color per molecule type) from a fixed number of molecule types, interacting through a network of catalytic reactions. **B)** GARD assemblies were simulated through growth-fission generations: assemblies grew by accretion of environmental molecules and underwent spontaneous fission upon reaching a critical size. Simulation then proceeded with one of the progeny assemblies until the total number of generations was exhausted. **C)** We detected self-replicators, for each run, according to the GARD model definition: self-replicators are clusters in molecular composition space that exhibit homeostatic growth, like attractors. **D)** We computed the $\Phi^r$ trajectory for each run, based on the relative compositions at every molecular step.

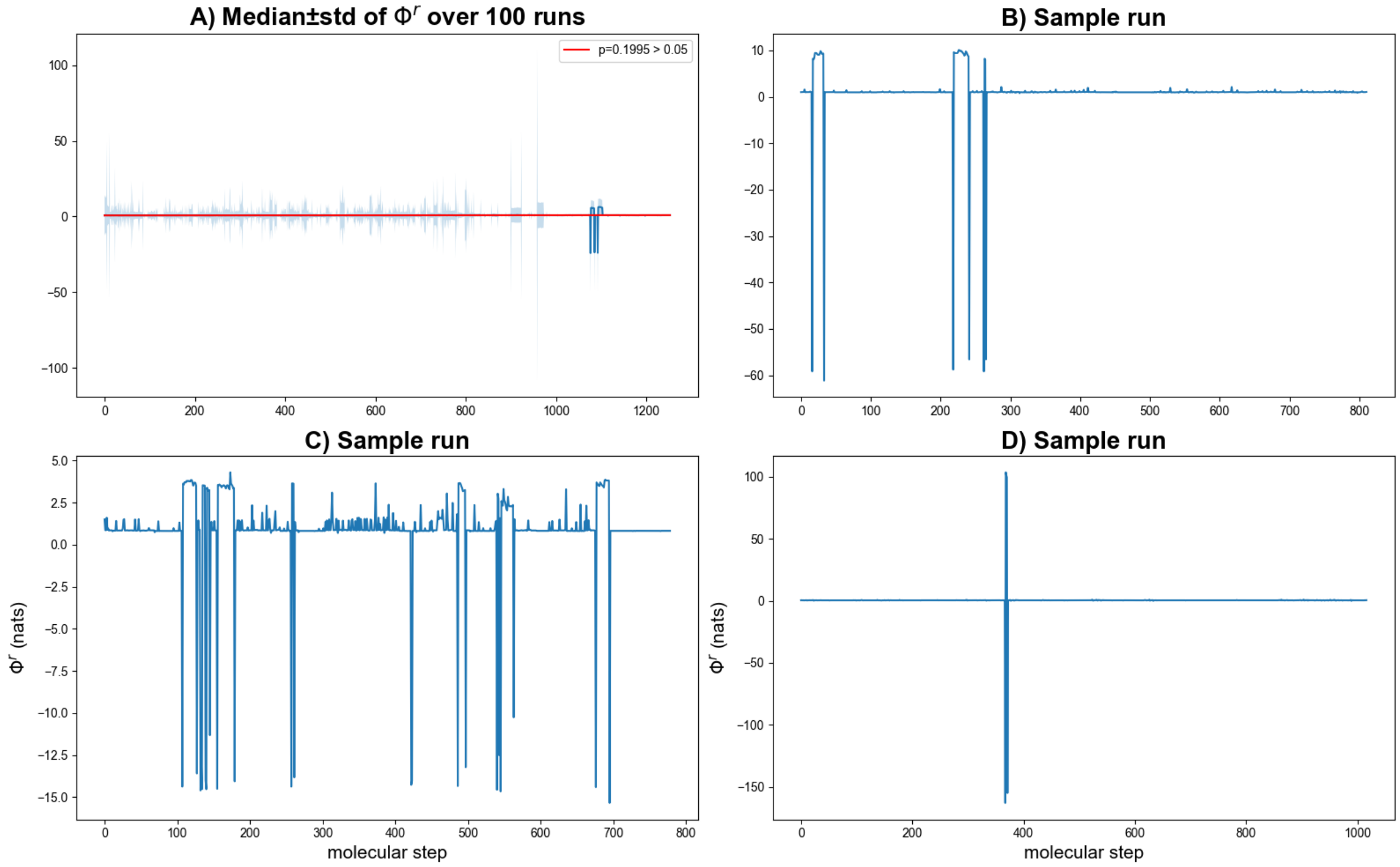

**Fig. 2. GARD prebiotic chemistries exhibit punctuated spiking in causal emergence.**
**A)** $\Phi^r$ (blue line, median ± std) across the 100 GARD runs, over molecular steps, with superimposed linear regression (red line) and its p-value. From this aggregate view, there is no significant trend. **B) C) D)** Sample $\Phi^r$ trajectories for individual runs. Despite the lack of an aggregate trend in panel A), individual runs exhibit punctuated spiking events in $\Phi^r$.

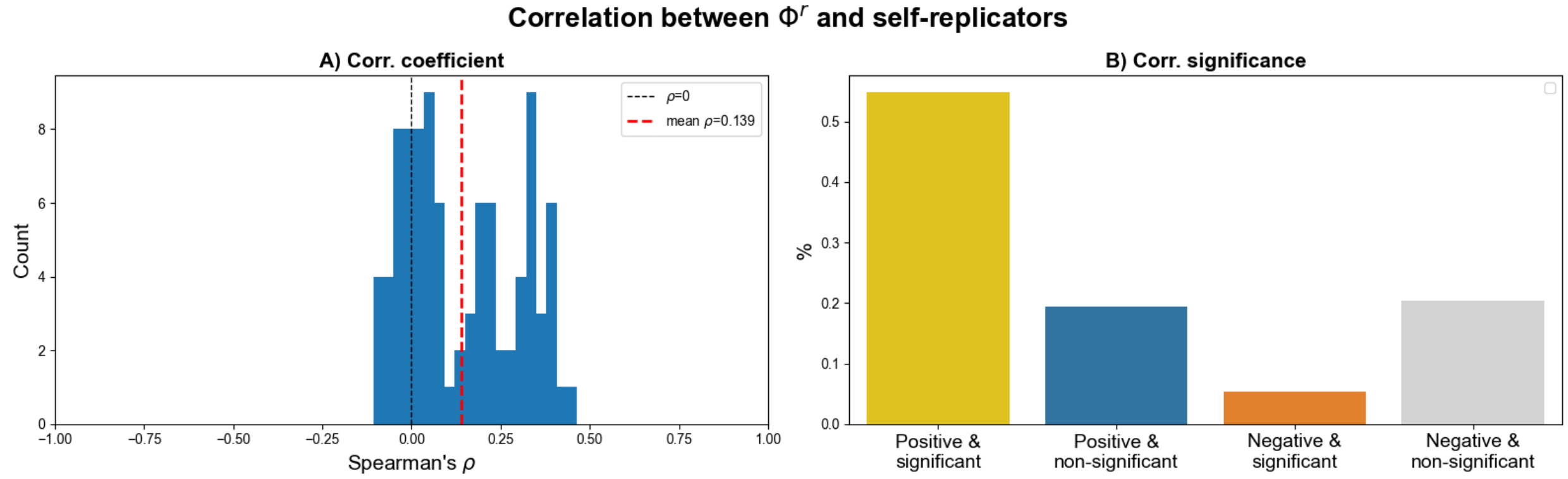

**Fig. 3. Occurrence of self-replicators is correlated with causal emergence.**
Spearman correlation analysis between changes in $\Phi^r$ and the occurrence of self-replication within each run. **A)** The correlation coefficients distribution for the 100 runs, with the positivity threshold 0 highlighted in the blue dashed line and the population means in the red dashed line. Results (with the one-sample t-test) showed that the mean is significantly positive. **B)** The 100 runs, broken down according to sign (pos. or neg.) and significance (sig. or not) of the correlation. There was a positive and significant correlation in most runs (54%), meaning that, typically, as $\Phi^r$ increased, so did the likelihood of settling in a self-replicating state.

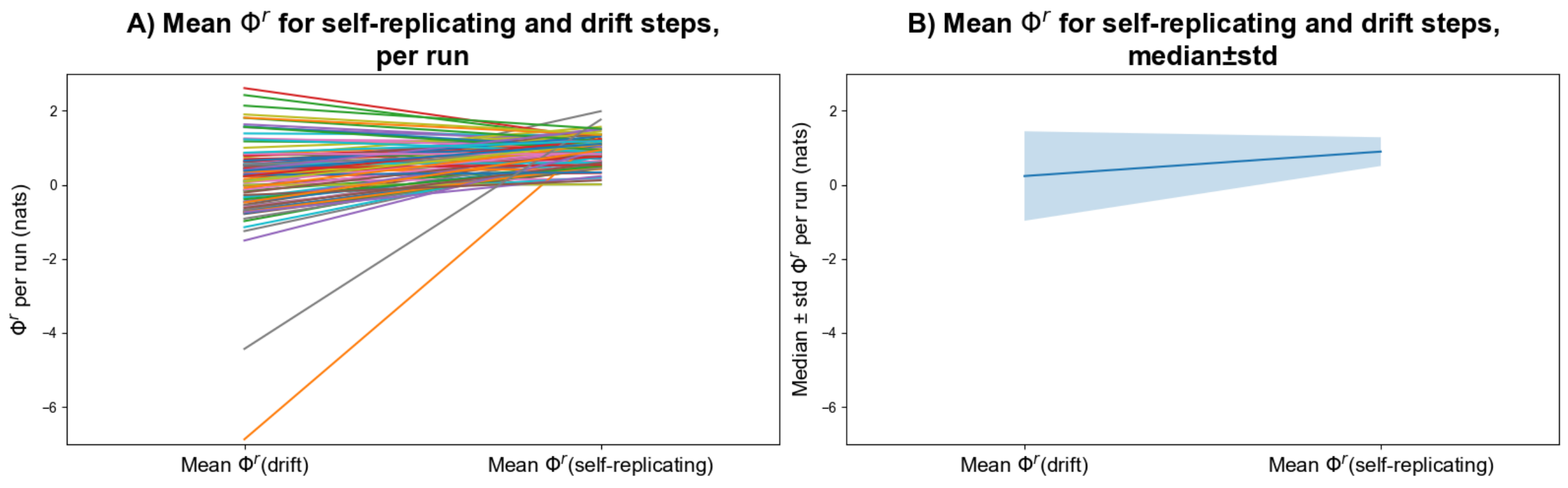

**Fig. 4. Causal emergence is higher in self-replicators than in non-self-replicators. A)** The change in mean $\Phi^r$ from the non-self-replicating (or *drift*) states (left points) to the self-replicating states (right points), one line per each of the 100 runs. **B)** The median ± std of the lines in B). Mean $\Phi^r$ was higher in self-replicating states in most runs (57%). Hence, self-replication corresponded to an increase in the causal emergence of the composomes.

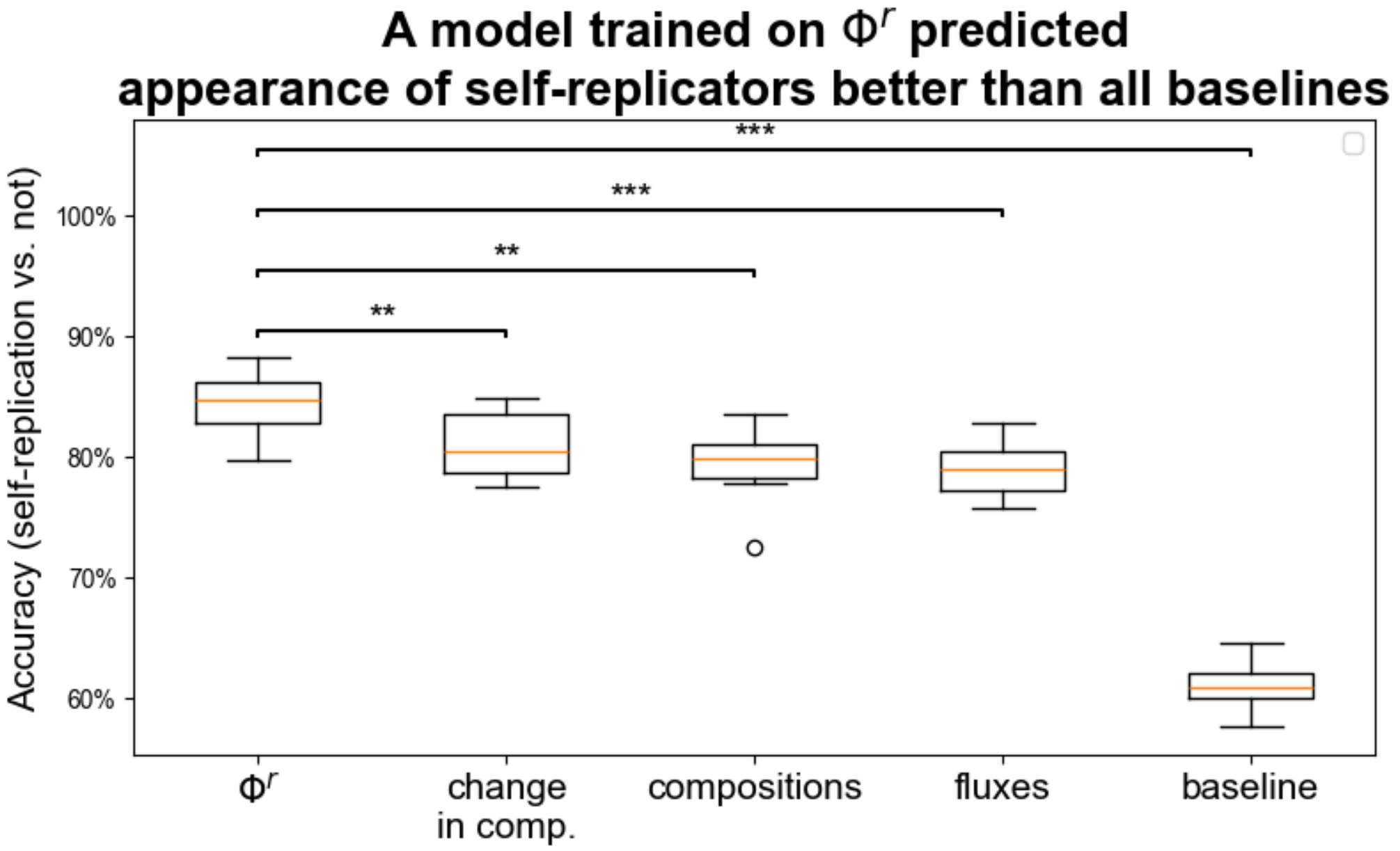

**Fig. 5. A machine learning model achieves superior prediction when predicting the probability of self-replicators from $\Phi^r$ instead of established metrics.**

Prediction performance (measured with binary accuracy) for four supervised learning models on the task of predicting the initial appearance of self-replicators (and when the assembly is not self-replicating) using $\Phi^r$ alone as input, or other metrics (change in composition, raw composition counts, fluxes between molecular species), or a dummy baseline that always predicts the most likely between self-replicator or not. The model trained on $\Phi^r$ achieved significantly better results than all the other models (p<0.01 at worst, Mann-Whitney).

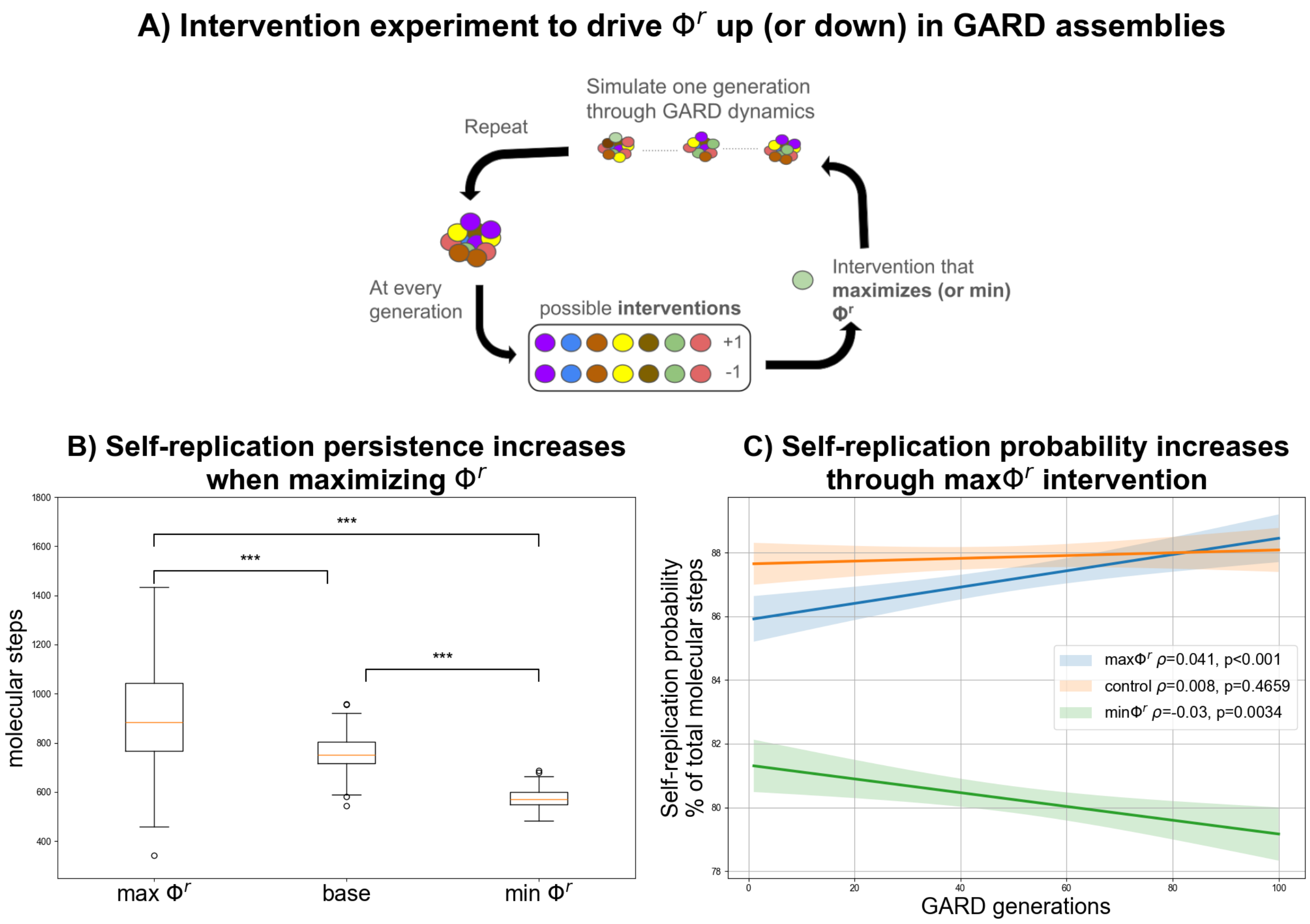

**Fig. 6. Intervening to drive $\Phi^r$ up increased self-replication persistence and probability of the entire medium.**

**A)** Intervention pipeline. At every generation right after fission, we add or delete the molecule that increased (or decreased) $\Phi^r$ the most, then let the system follow the natural GARD dynamics until the next generation. **B)** Distribution of the self-replication persistence (in molecular steps) with interventions to maximize $\Phi^r$ (max$\Phi^r$), minimize $\Phi^r$ (min$\Phi^r$), or neither (control). The differences are significant with the Mann-Whitney test. **C)** For the three treatments, the regression lines of the self-replication probability (% molecular steps over the total) over the GARD generations, with the 95% confidence interval. Not only did interventions that drove $\Phi^r$ up result in more persistent self-replication, but the effect on self-replication probability accumulated over time as more interventions were applied. Crucially, interventions that minimized $\Phi^r$ had the opposite effect. Hence, causal emergence causally modulated self-replication.

| | Persistence | Probability | Consistency | Time to first replicator |
|---|---|---|---|---|
| max$\Phi^r$ | 874±233 | 88±3% | 0.52±0.04 | 36±26% |
| control | 716±198 | 88±3% | 0.38±0.06 | 37±27% |
| min$\Phi^r$ | 559±99 | 80±3% | 0.42±0.04 | 40±28% |

**Table 1. Properties of self-replicators in GARD assemblies, by three treatments: with interventions to maximize $\Phi^r$ (max$\Phi^r$), minimize $\Phi^r$ (min$\Phi^r$), or neither (control).**

Persistence measures the total lifetime of self-replicators in molecular steps; the probability is the fraction of steps spent in self-replication; consistency is the Pearson's $\rho$ between consecutive steps (the higher, the more compact and less erratic the self-replicators are); while the time to first replicator is the number of molecular steps before the appearance of the first replicator. The max$\Phi^r$ intervention achieved higher persistence and consistency, while being comparable to the control in terms of probability and time to first replicator. Conversely, the min$\Phi^r$ obtained lower persistence and probability.